\documentclass[runningheads]{llncs}

\usepackage[english]{babel}

\usepackage{diagbox}
\usepackage{booktabs}

\usepackage{listings}
\usepackage{cite}
\usepackage{amsmath,amssymb,amsfonts}
\usepackage{multirow}
\usepackage{algorithmic}
\usepackage{graphicx}
\usepackage{textcomp}
\usepackage[utf8x]{inputenc}
\usepackage[table,xcdraw]{xcolor}
\usepackage[pdfa]{hyperref}
\usepackage{csquotes}
\usepackage[english]{babel}
\usepackage{algorithmic}
\usepackage{graphicx}
\usepackage{textcomp}
\usepackage{paralist}
\usepackage{url}

\begin{document}
\title{Deep Feature Learning for Medical Acoustics}

\author{Alessandro Maria Poirè \and
  Federico Simonetta\orcidID{0000-0002-5928-9836} \and
  Stavros Ntalampiras\orcidID{0000-0003-3482-9215}}
\authorrunning{A. Poirè et al.}
%
\institute{LIM -- Music Informatics Laboratory\\
  Department of Computer Science\\
  University of Milano\\
  \email{stavros.ntalampiras@unimi.it} \\ \url{https://www.lim.di.unimi.it/}}
%

\maketitle

\begin{abstract}

  The purpose of this paper is to compare different learnable frontends in
  medical acoustics tasks. A framework has been implemented to classify human
  respiratory sounds and heartbeats in two categories, i.e.\ healthy or affected
  by pathologies. After obtaining two suitable datasets, we proceeded to
  classify the sounds using two learnable state-of-art frontends -- LEAF and
  nnAudio -- plus a non-learnable baseline frontend, i.e. Mel-filterbanks. The
  computed features are then fed into two different CNN models, namely VGG16 and
  EfficientNet. The frontends are carefully benchmarked in terms of the number
  of parameters, computational resources, and effectiveness.

  This work demonstrates how the integration of learnable frontends in neural
  audio classification systems may improve performance, especially in the field
  of medical acoustics. However, the usage of such frameworks makes the needed
  amount of data even larger. Consequently, they are useful if the amount of
  data available for training is adequately large to assist the feature learning
  process.

\end{abstract}


\section{Introduction}
  Cardiovascular and respiratory diseases are the leading cause of mortality
  worldwide; it is estimated that in 2019 17.9 million people died due to
  cardiovascular diseases, representing the first and second cause of death
  worldwide (32 \% of all deaths worldwide), followed by respiratory
  disease~\cite{who-heart,who-resp}. Therefore, considerable efforts have been
  devoted to research for the improvement of the early diagnosis and routine
  monitoring of patients with cardiovascular and respiratory diseases. A large
  portion of the research has focused on the auscultation of respiratory sounds
  and heart tones. Indeed, these diseases, such as asthma, COPD, pneumonia, heart
  murmurs, heart valve abnormalities, and arrhythmia, are associated with
  distinct sound patterns. Such abnormal breathing sounds in the lungs are called
  adventitious sounds~\cite{who-heart}. A similar phenomenon can be observed relatively to abnormal blood flows in the   heart, which can also cause characteristics noises.

  To the purpose of screening these cardiovascular and respiratory diseases,
  cardiac auscultation by phonocardiograms (PCG) and pulmonary auscultation are
  among the most important tools. The auscultation happens via an electronic
  stethoscope capable of digitally recording PCGs and respiratory sounds.
  However, this process is based on the availability of an expert as well as on
  his degree of competence. Thus, the need to automate the diagnosis process has
  arisen in recent years, bringing the development of algorithms able of
  classifying heart or pulmonary sounds. Such algorithms are usually based on
  Machine Learning technologies with the aim of assisting physicians in the
  diagnosis of health diseases, as well as providing patients with effective
  auto-diagnosis tools where physicians are not available \cite{9159888,Ntalampiras2020}.

  Artificial Neural Networks (ANN) comprise the most used approach for the
  classification of heart and pulmonary sounds~\cite{physionet-cuore}. ANNs
  require discriminating features of the signal as input; usually such features
  are time-frequency representations of an audio signal, such as spectrograms, Mel
  spectrograms and Mel-frequency cepstral coefficient
  (MFCC)~\cite{features1,features2,features3}. Recent studies, regarding sound classification
  in general, show that using Log-Mel spectrograms has significant improvements
  on the efficiency of the neural network~\cite{logmel,Purwins2019}. Some studies
  also adopted Wavelet-based representations, but such features were only little
  explored compared to FFT-based ones~\cite{wavelet}.


  The goal of this work is the comparison of various frontends, i.e.
  feature-extraction methods. Indeed, various frontends for neural features
  extraction were recently proposed in the field of audio signal processing.
  Specifically, two learnable frontends for audio processing received a large
  attention -- LEAF and nnAudio~\cite{leaf,nnaudio}. Both the two frontends
  allow to compute time-frequency representations specifically crafted for the
  learning problem. This study aims at assessing if the learned features can
  improve the efficiency of ANN for Medical Acoustics and therefore we compare
  them to a standard representation method based on Log-Mel-spectrograms.



  The contributions of this work are:
  \begin{itemize}
    \item a binary classification method for respiratory and heartbeat sounds;
    \item comparison of LEAF and nnAudio frameworks with traditional hand-crafted
          features for audio processing;
    \item efficiency and effectiveness benchmarks of different feature extraction
          strategies using different types of Neural Networks;
  \end{itemize}

  To the sake of reproducibility, the source-code used for this work is fully available
  online~\footnote{\url{https://github.com/LIMUNIMI/Feature-Learning-Medical-Acoustics}}.


\section{The Considered Frontends}
  \label{sec:frontends}
  
  \begin{figure}[t]
    \centering
    \includegraphics[width=0.75\textwidth]{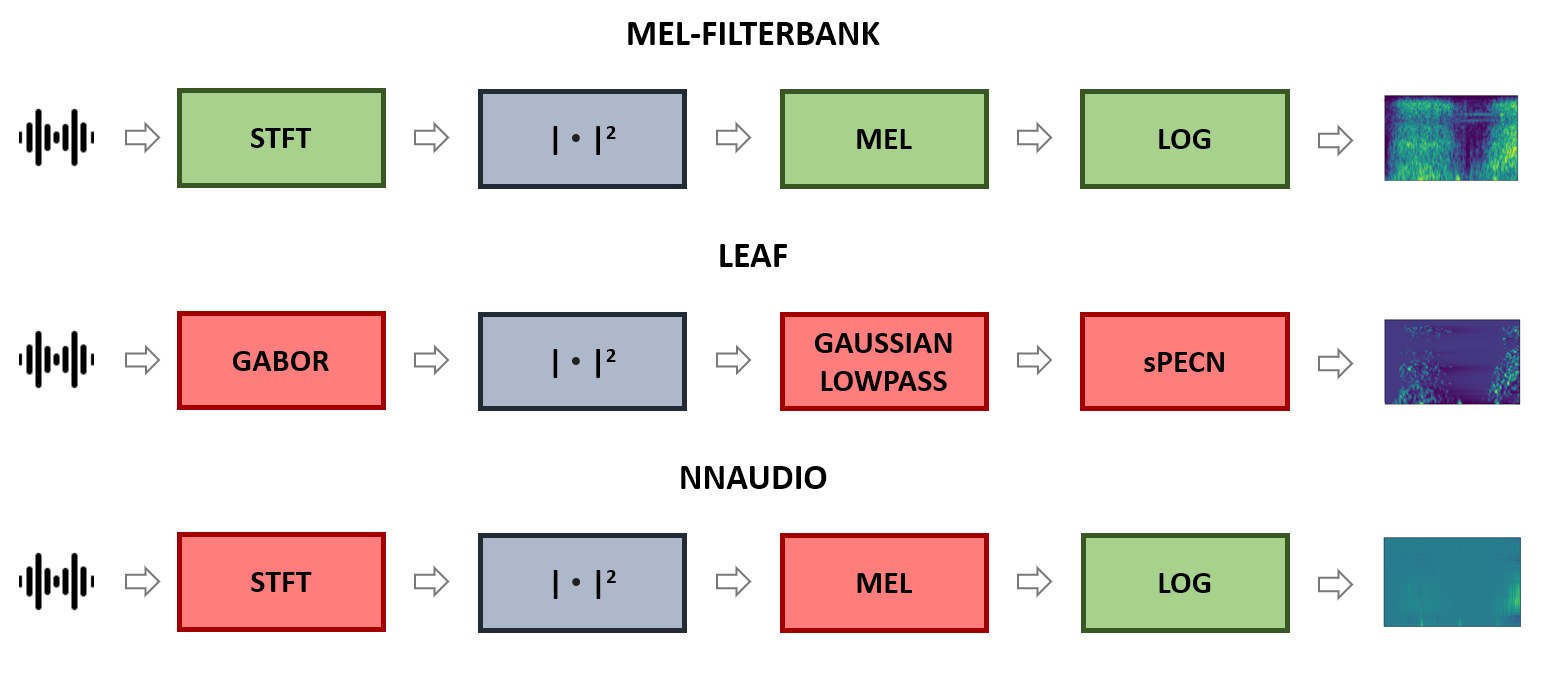}
    \caption{Breakdown of Mel-filterbanks, LEAF and nnAudio frontends. Orange boxes are fixed, while computations in blue boxes are learnable. Grey boxes represent activation functions.}
    \label{fig:breakdown}
  \end{figure}

  As mentioned above, the purpose of this paper is to compare  two
  learnable frontends, LEAF and nnAudio.	LEAF and nnAudio are features
  extractors that, unlike Mel-filterbank, are completely trainable during the
  neural network training process. Interestingly, all audio features extraction operations, such
  as filtering, pooling, compression and normalization are learnable.

  Log-Mel spectrograms are the most used time-frequency representation for neural
  classification tasks in the field of medical acoustics~\cite{reason1,reason2,reason3,reason4,reason5};
  it is for this reason that they have been chosen as a baseline for comparing
  representations produced by learnable frontends.

  The three frontends are depicted in Fig.~\ref{fig:breakdown}.

  \subsection{Mel-filterbanks}

    Mel-filterbank is a fixed frontend that receives waveforms as an input, and
    produces Log-Mel spectrograms as output. It is fixed because the parameters
    that control it are non-learnable, that is, they do not change during the
    training of the network.

    A Mel-filterbank is applied to an audio excerpt to obtain the Log-Mel
    spectrograms. More specifically, we first compute the spectrogram of the audio
    excerpt using the squared modulus of the short-term Fourier transform (STFT).
    Then, the spectrogram is passed through a bank of triangular bandpass filters,
    spaced logarithmically according to the Mel scale. The Mel scale is designed to
    replicate the non-linearity of human pitch perception. Finally, to reflect the
    non-linearity of the human loudness sensitivity, the resulting coefficients are
    passed through a logarithmic compression.

    Log-Mel spectrograms have various parameters that should be finely tuned,
    adding a large number of hyper-parameters to the resulting pipeline and making
    the designing of Machine Learning methods more complex.

  \subsection{LEAF}

    LEAF (LEarnable Audio Frontend)~\cite{leaf} is a neural
    network-based frontend to extract features such as Mel spectrograms. Being a
    neural network, it can be trained inside any neural architecture to discover
    task-specific features, adding only a few parameters to the model. This
    frontend learns all operations of audio features extraction, from filtering to
    pooling, compression and normalization.

    In the first stage of filtering the sound wave passed through a bank of Gabor
    bandpass filters followed by a non-linearity. Then, the temporal resolution of
    the signal is reduced in the ``pooling'' phase. Finally, the dynamic range is
    optimized with a compression and/or normalization stage.

  \subsection{nnAudio}

    nnAudio (neural network Audio)~\cite{nnaudio} is a neural network
    based frontend able to extract Mel spectrograms as features. nnAudio uses
    convolutional neural networks to perform the conversion from time domain to
    frequency domain, and it can be trained together with any classifier.

    As input, nnAudio receives a waveform from which it extracts the
    Mel-spectrogram via a learnable process. The frontend first computes the STFT
    using a Convolutional Neural Network, and then applies a bank of Mel filters.
    The values of the Mel filter bank are used to initialize the weights of a
    single-layer fully-connected neural network. Each time step of the STFT is sent
    in this fully-connected layer initialized with Mel weights. The Mel filter bank
    therefore must only be created during the initialization of the neural network.
    All of these weights are trainable.

\section{Models}
  \label{sec:models}

  To test the frontends under different conditions, two different well-known CNNs
  were chosen for the classification phase: \textit{EfficientNet-B0} and
  \textit{VGG16}~\cite{efficientnet,vgg16}.

  \subsection{EfficientNet}
    The EfficientNet models are a family of artificial neural networks where the
    basic building block is the Mobile Inverted Bottleneck Conv Block (MBConv). The
    Efficient-Net family includes 8 models (from B0 to B7): as the number
    increases, the complexity of the network increases. The main idea of
    EfficientNet is to start from one simple, compact and computationally efficient
    structure, and gradually increasing its complexity. Unlike other CNN models,
    EfficientNet uses a new activation feature known as Swish, rather than the
    classic ReLU function. The “lightweight” version of EfficientNet
    (EfficientNetB0, with $\sim$4M parameters) has been adopted as
    first classifier.

  \subsection{VGG}
    The VGG16 version of VGG was adopted as another classifier. VGG stands for
    Visual Geometry Group; It is a standard multi-level CNN architecture. According
    to the number of layers, the various versions of VGG are named, for example
    VGG11 has 11 layers, VGG16 has 16 layers, VGG19 has 19 layers and so on.

    VGG16 is a deep 16-layer neural network; this means that it is quite large, and
    has a total of about 138 million parameters. However, its architecture is
    relatively simple and straightforward.


\section{Datasets}
  \label{sec:datasets}

  \begin{table}[t]
    \caption{\label{tab:datasets}Statistics on the datasets used}
    \centering
    \begin{tabular}{cc|c|c|}
      \cline{3-4}
      &  & \begin{tabular}[c]{@{}c@{}}\textbf{Respiratory Sound} \\ \textbf{Database}~\cite{icbhi_dataset}\end{tabular} & \begin{tabular}[c]{@{}c@{}}\textbf{Heartbeat Physio-Net} \\ \textbf{Database}~\cite{physionet_dataset}\end{tabular} \\ \hline
      \multicolumn{1}{|c|}{\multirow{4}{*}{\textbf{Normal Samples}}}   & Train set      & 2732                       & 20461                         \\ \cline{2-4} 
      \multicolumn{1}{|c|}{}                                  & Validation set & 546                        & 4092                          \\ \cline{2-4} 
      \multicolumn{1}{|c|}{}                                  & Test set       & 364                        & 2728                          \\ \cline{2-4} 
      \multicolumn{1}{|c|}{}                                  & \textbf{Total}          & \textbf{3642}                       & \textbf{27281}                         \\ \hline
      \multicolumn{1}{|c|}{\multirow{4}{*}{\textbf{Abnormal Samples}}} & Train set      & 2442                       & 5284                          \\ \cline{2-4} 
      \multicolumn{1}{|c|}{}                                  & Validation set & 488                        & 1057                          \\ \cline{2-4} 
      \multicolumn{1}{|c|}{}                                  & Test set       & 326                        & 704                           \\ \cline{2-4} 
      \multicolumn{1}{|c|}{}                                  & \textbf{Total}          & \textbf{3256}                       & \textbf{7045}                          \\ \hline
      \multicolumn{2}{|c|}{\textbf{Total}}                                              & \textbf{6898}                       & \textbf{34326}                         \\ \hline
    \end{tabular}
  \end{table}

  In order to compare the proposed frontends, tests were performed for two
  different medical acoustics tasks: anomaly detection in respiratory sounds and
  in heartbeat recordings. The datasets differ in content and quantity of
  elements, so that frontends can be tested under different conditions.
  
  Table~\ref{tab:datasets} shows summary statistics about the used datasets.

  \subsection{Respiratory dataset}
    \label{sec:ICBHI}
    The first database is the Respiratory Sound database~\cite{icbhi_dataset},
    created to support the scientific challenge organized at the International
    Conference on Biomedical Health Informatics - ICBHI
    2017~\cite{icbhi}.

    The database consists of a total of 5.5 hours of records containing 6898
    respiratory cycles, of which 1864 contain crackles, 886 contain wheezes and 506
    contain both crackles and wheezes.

    The total number of audio samples was 920, obtained from 126 participants. The
    recordings were collected using heterogeneous equipment and their duration
    ranged from 10 to 90 seconds. For each audio recording, the time-mark list of
    start and end time of each respiratory cycle is provided. The sampling
    frequency of the recordings varies, with values of 4 kHz, 10 kHz or 44.1 kHz;
    in the preprocessing phase they are resampled at 4kHz. It is currently the
    largest publicly available respiratory sound database.

    The level of noises -- cough, speech, heartbeat, etc. -- in some breathing
    cycles is relatively high representing real-life conditions very well. Respiratory cycles
    were noted by experts, dividing them into four categories: crackles, wheezes, a
    combination of them, or no adventitious sounds.

    In this work we have chosen to use only two labels: normal and abnormal. The
    normal class covers sounds categorized as non-adventitious, while the abnormal
    class includes sounds containing crackles, wheezes, or a combination of them.
    In this way, the two resulting classes are more balanced, accounting 3642
    normal sounds and 3256 abnormal cycles.

  \subsection{Heartbeat dataset}
    \label{sec:CinC}
    The second database is the Heartbeat Physio-Net
    Database~\cite{physionet_dataset}, created specifically for the 2016 PhysioNet
    Computing in Cardiology (CinC) Challenge~\cite{physionet}.

    This database contains a total of 3153 heart sound recordings from 764 healthy
    and pathological patients. The recordings have a duration from 5 to 120
    seconds, obtaining about 25 hours of sound material. All audio samples were
    recorded with a sampling rate of 2kHz or resampled to the same rate. In the
    database there are recordings labeled as unsure, i.e. with a very low signal
    quality. These audio samples were omitted from the test, leaving a total of
    2872 recordings for the training, validation and testing phases. All
    phonocardiograms in the database are categorized into two types: normal and
    abnormal. Recordings with the normal label come from healthy patients, while
    those with the abnormal label come from patients with pathologies such as
    coronary artery disease and heart valve defects (mitral valve prolapse
    syndrome, mitral regurgitation, aortic regurgitation, aortic stenosis and valve
    surgery).

    As in the Respiratory Database, the data includes not only ``clean'' heart
    sounds, but also very noisy recordings, providing an accurate representation of
    real life conditions.

\section{Experiments}

  \begin{figure*}[t]
    \centering
    \includegraphics[width=0.75\textwidth]{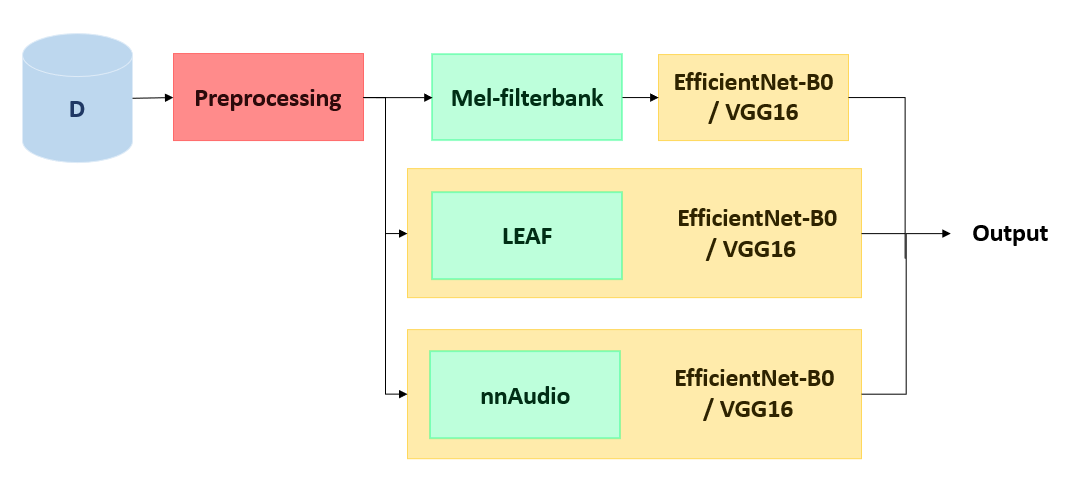}
    \caption{General scheme of the neural network architecture.}
    \label{fig:experiments}
  \end{figure*}

  This section describes the experiments performed to compare the various
  frontends and models described in
  Sections~\ref{sec:frontends}~and~\ref{sec:models}. The generic
  workflow is shown in Fig.~\ref{fig:experiments}.

  \subsection{Pre-processing}

    Pre-processing included segmentation, filtering, resampling, and padding.

    For the respiratory database, the samples were segmented following the
    time-marks annotations indicating each respiratory cycle -- see
    Sec.~\ref{sec:ICBHI}. For instance, the audio file
    \texttt{107\_2b5\_Pr\_mc\_AKGC417L.wav} is 8.97 seconds long; it has been segmented according
    to the indicated time-marks thus producing 4 audio files:
    \texttt{file\_1.wav} (from 0.077 to 1.411 of the original file), \texttt{file\_2.wav} (from 1.411 to 3.863), \texttt{file\_3.wav} (from 3.863 to 6.601) and
    \texttt{file\_4.wav} (from 6.601 to 8.97).

    For the heartbeat database, instead, the individual audio files of variable
    length (from 6 to 120 seconds) were segmented into 2-second files. For
    instance, a file initially lasting 10 seconds is split into 5 files of 2
    seconds each. In this way, the total number of samples becomes 34326.

    Subsequently, the audio files obtained from the segmentation phase are filtered
    through a 12th order Butterworth band-pass filter, with cut-off
    frequencies [120 - 1800] Hz for the respiratory database,
    and [25, 400] Hz for the heartbeat database. This eliminates
    the components of sound caused by coughing, intestinal noises, stethoscope
    movement and speech.

    All audio files were then resampled at 4 KHz (only in the Respiratory dataset,
    the Heartbeat dataset was already sampled at 4kHz) and truncated or zero-padded
    so that they lasted exactly 2 seconds.

    \begin{figure}[t]
        \centering
        \includegraphics[width=\textwidth]{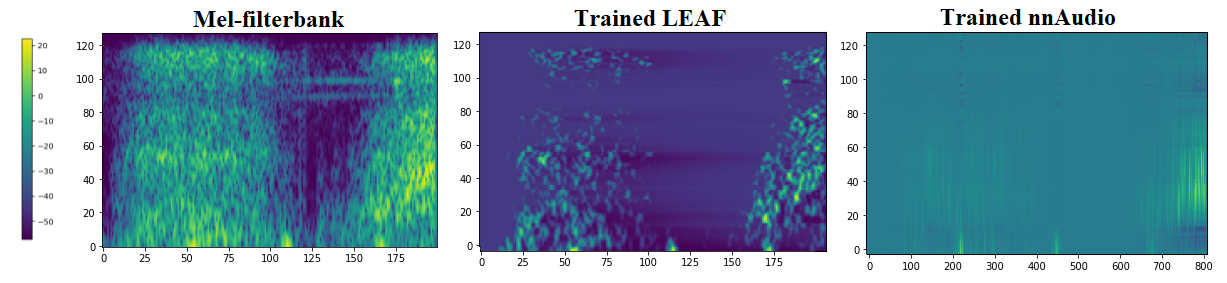}
        \caption{Features learned on a sample from the Respiratory Sound Database~\cite{icbhi_dataset} by LEAF and nnAudio compared with classical Mel-spectrograms.}
        \label{fig:resp_feat}
    \end{figure}

  \subsection{System parameterization}
    To better compare the frontends considered, the same hyper-parameters were used
    in the tests.

    In all three frontends, after various experiments, it was decided to use a
    window length of 30ms, with a window stride of 10ms. The frequency range of the
    Mel filters was set at [100, 2000] Hz in the tests with
    respiratory dataset, and [25, 1000] Hz in the tests with
    heratbeat dataset. 128 Mel-filters were used. Only in the LEAF frontend some
    parameters have been changed in respect to their factory defaut values;
    specifically, in the PCEN compression layer, the alpha and root parameters that
    control the amount of compression applied, respectively alpha = 2, root = 4.

    In the VGG16 classifier, two dropout layers with a value of 0.5 have been added
    between the last two fully connected layers. The dropout layer prevents the
    co-adaptation of a neural network, disabling some nodes of the network during
    the training phase with a specific probability (0.5 in this case). In
    EfficientNetB0 the drop connect rate parameter has also been changed, setting
    it to 0.5.

    The training was carried out considering a period of 300 epochs for the
    heartbeat test, and 200 epochs for the respiratory test. The batch size was set
    at 64, while the learning rate as set at 1e-5, and ADAM was chosen as
    weight-update algorithm.

    We empirically found an optimal split size using 75\% of the dataset for the
    train set, 15\% for the validation set, and 10\% for the test set.
    
\begin{figure}[t]
        \centering
        \includegraphics[width=\textwidth]{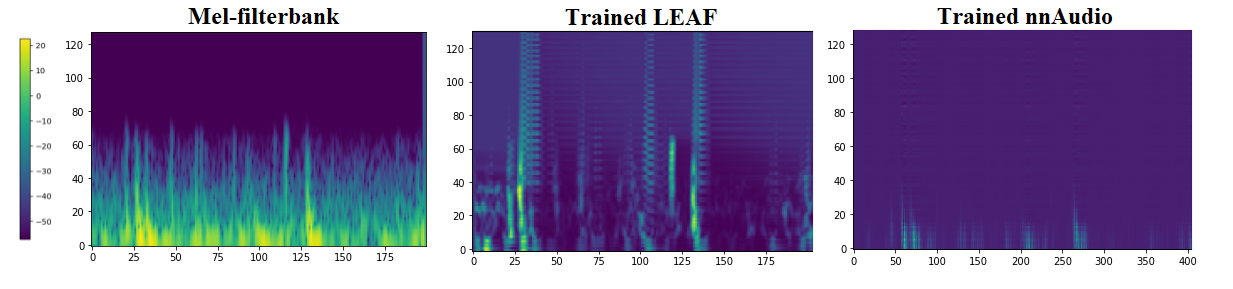}
        \caption{Features learned on a sample from the Heartbeat Physio-Net Database~\cite{physionet_dataset} by LEAF and nnAudio compared with classical Mel-spectrograms.}
        \label{fig:heart_feat}
    \end{figure}

    Examples of features learned on the two datasets are shown in Figures~\ref{fig:resp_feat}~and~\ref{fig:heart_feat}.

\section{Results}

    \begin{table}[b]
      \centering
      \caption{
      \label{tab:pvals}
      McNemar p-values corrected with Bonferroni-Holm method.}
      \scalebox{1}{
        \begin{tabular}{c|c|c|c}
          \toprule
                                       & \textbf{\textit{Mel-LEAF}} & \textbf{\textit{Mel-nnAudio}} & \textbf{\textit{LEAF-nnAudio}} \\
          \midrule
          \textbf{\textit{Resp. Eff.}} & 0.2010                     & 0.0500                        & 0.5263                         \\
          \midrule
          \textbf{\textit{Resp. VGG}}  & 0.3998                     & 0.0366                        & 0.3998                            \\
          \midrule
          \textbf{\textit{Heart. Eff.}}
                                       & 9.8515e-07                 & 0.4194                        & 2.6648e-05                     \\
          \midrule
          \textbf{\textit{Heart. VGG}}
                                       & 1.1219e-03                 & 9.1734e-09                    & 7.1666e-03                     \\

          \bottomrule
        \end{tabular}}

    \end{table}

  In order to compare the proposed frontends, two tests with two different
  datasets were formulated: ``Test 1 - Respiratory'' and ``Test 2 - Heartbeat''.
  The datasets differ in content and quantity of elements, so the frontends can
  be tested under different conditions.

  The evaluation metrics used in this study are balanced accuracy, True Positive
  Rate (TPR) and True Negative Rate (TNR):
  \begin{equation}
    \begin{split}
      \begin{gathered}
        Balanced Accuracy = \frac{TPR + TNR}{2} \\
        TPR = \frac{True Positive}{True Positive + False Negative} \\
        TNR = \frac{True Negative}{True Negative + False Positive}
        \label{metricEQ}
      \end{gathered}
    \end{split}
  \end{equation}

  \subsection{Test 1 - Respiratory}

    As shown in Tables~\ref{tab:res_resp_vgg}~and~\ref{tab:res_resp_eff}, VGG16
    always achieves better results than EfficientNet. However, using VGG16, the
    differences between the three frontends are larger, acheiving 6\% of difference
    in accuracy between nnAudio and Mel-filterbank.

    Surprisingly, we found that with VGG16 the baseline method outperforms the
    learnable frontends, proving the well-design of old Log-Mel spectrograms
    compared to newer neural network frameworks.

    When EfficientNet is used, instead, a small difference emerges that awards
    the learnable frontends, especially nnAudio; however, McNemar test with
    Bonferroni-Holm correction finds no statistically significant difference among
    the three.

    Specific p-values are shown in Table~\ref{tab:pvals}.

    \begin{table}[b]
      \caption{
      \label{tab:res_resp_vgg}
      Comparison results using VGG16 as classifier  on the ICBHI dataset~\cite{icbhi_dataset}. Only Mel-filterbank and nnAudio accuracies show a statistical significance ($p\sim0.04$ using McNemar with Bonferroni-Holm correction).}
      \centering
      \scalebox{1}{
        \begin{tabular}{|c|c|c|c|}
          \hline
          \%                               & \textbf{\textit{Balanced Accuracy}} & \textbf{\textit{TPR}} & \textbf{\textit{TNR}} \\
          \hline
          \textbf{\textit{Mel-filterbank}} & \textbf{80.21}                      & \textbf{81.42}        & \textbf{79.01}        \\
          \hline
          \textbf{\textit{LEAF}}
                                           & 77.47                               & 80.87                 & 74.07                 \\
          \hline
          \textbf{\textit{nnAudio}}        & 74.12                               & 77.86                 & 70.37                 \\
          \hline
        \end{tabular}}

    \end{table}

    \begin{table}[t]
      \caption{
      \label{tab:res_resp_eff}
      Comparison results using EfficientNet-B0 as classifier on the ICBHI dataset~\cite{icbhi_dataset}. Only Mel-filterbank and nnAudio accuracies show a small significance ($p\sim0.05$ using McNemar with Bonferroni-Holm correction).}
      \centering
      \scalebox{1}{
        \begin{tabular}{|c|c|c|c|}
          \hline
          \%                               & \textbf{\textit{Balanced Accuracy}} & \textbf{\textit{TPR}} & \textbf{\textit{TNR}} \\
          \hline
          \textbf{\textit{Mel-filterbank}} & 61.05                               & 62.84                 & \textbf{59.26}        \\
          \hline
          \textbf{\textit{LEAF}}
                                           & 61.40                               & 66.94                 & 55.86                 \\
          \hline
          \textbf{\textit{nnAudio}}        & \textbf{61.74}                      & \textbf{68.85}        & 54.63                 \\
          \hline
        \end{tabular}}

    \end{table}

  \subsection{Test 2 - Heartbeat}

    Even in this scenario VGG16 was better than EfficientNet in all the tests.
    Nevertheless and contrarily to the respiratory task, the Mel-filterbank was
    surpassed by both nnAudio and LEAF.

    Tables~\ref{tab:res_heart_vgg}~and~\ref{tab:res_heart_eff} show that LEAF
    achieves the better accuracy using both VGG16 and EfficientNet. However, when using EfficientNet, the
    best TNR was achieved by nnAudio. Note that TNR is
    particularly important in first-screening diagnosis, because low TNR is
    associated with a high false negative rate, meaning that false negatives are
    common. When a false negative prediction happens, the therapeutic intervention
    may be delayed with catastrophic consequences.

    Specific p-values are shown in Table~\ref{tab:pvals}.

    \begin{table}[t]
      \caption{
      \label{tab:res_heart_vgg}
      Comparison results using VGG16 as classifier on the heart beats dataset~\cite{physionet_dataset}. All p-values between accuracies are $<<0.05$ (McNemar with Bonferroni-Holm correction)}

      \centering
      \scalebox{1}{
        \begin{tabular}{|c|c|c|c|}
          \hline
          \%                               & \textbf{\textit{Balanced Accuracy}} & \textbf{\textit{TPR}} & \textbf{\textit{TNR}} \\
          \hline
          \textbf{\textit{Mel-filterbank}} & 90.71                               & \textbf{96.22}        & 85.20                 \\
          \hline
          \textbf{\textit{LEAF}}
                                           & \textbf{92.29}                      & 95.29                 & \textbf{89.30}        \\
          \hline
          \textbf{\textit{nnAudio}}        & 91.41                               & 93.64                 & 89.16                 \\
          \hline
        \end{tabular}}

    \end{table}
    \begin{table}[t]
      \caption{
      \label{tab:res_heart_eff}
      Comparison results using EfficientNet-B0 as classifier on the heart beats dataset~\cite{physionet_dataset}. All comparisons of accuracies revealed stastical significance except between Mel-filterbank and nnAudio (McNemar test with Bonferroni-Holm correction).}
      \centering
      \scalebox{1}{
        \begin{tabular}{|c|c|c|c|}
          \hline
          \%                               & \textbf{\textit{Balanced Accuracy}} & \textbf{\textit{TPR}} & \textbf{\textit{TNR}} \\
          \hline
          \textbf{\textit{Mel-filterbank}} & 81.12                               & 90.92                 & 71.33                 \\
          \hline
          \textbf{\textit{LEAF}}
                                           & \textbf{84.36}                      & \textbf{95.40}        & 73.31                 \\
          \hline
          \textbf{\textit{nnAudio}}        & 83.51                               & 92.52                 & \textbf{74.50}        \\
          \hline
        \end{tabular}}
    \end{table}
    
  \subsection{Overall}

    Comparing the results obtained from the two tests Test 1 and Test 2, we note
    that the best scores were achieved in Test 2 (Heartbeat), with the LEAF
    frontend. We theorize that LEAF performed better in Test 2 than Test 1 due to
    the size of the phonocardiogram database, which is much larger than the
    respiratory sound database.

    Moreover, the different balancing of the two databases is probably the reason
    why TPR and TNR are more distant in Test 2 than in Test 1. Indeed, the
    Respiratory database has the most balanced classes compared to the Heartbeat
    database -- see Table~\ref{tab:datasets}.
    
    In general, the 3 frontends learn different features, as shown in Figures~\ref{fig:resp_feat}~and~\ref{fig:heart_feat}. Namely, nnAudio learns more sparse representations that focus on low frequencies. On the contrary, LEAF learns representations less sparse and well distributed across the frequency space. Compared with classical Mel-filterbanks, both of them seems to learn specific characteristics that are relevant for the classification. We theorize that LEAF learned features work by extracting discriminative local descriptors in the time-frequency space similarly to audio fingerprint algorithms~\cite{Wang2003AnIS,Ke2005ComputerVF}. nnAudio, instead, extracts blurred regions that are likely less characteristics of the single excerpt. Moreover, LEAF manages to handle both positive and negative values, while nnAudio's activation functions only return non-negative values, thus deleting possibly useful information.

\section{Conclusion}

  This work has shown how the integration of learnable frontends in
  classification systems with convolutional neural networks can improve results
  in the field of medical acoustics. The tests carried out show that learnable
  frontends are particularly useful when there is a sufficient amount of 
  available data (Test 2), while using small data-sets (Test 1) prevent them
  from learning accurate features to surpass the classic hand-crafted methods.

  The proposed method therefore stands as a valid alternative to traditional
  feature extraction methods as long as they are used in contexts with a large
  amount of data available.

  \bibliographystyle{splncs04}
  \bibliography{IEEEabrv,bibliography}

\end{document}